\documentclass[conference]{IEEEtran}
\usepackage[letterpaper, left=0.68in, right=0.68in, top=0.75in, bottom=1in]{geometry}
\IEEEoverridecommandlockouts
\usepackage[T1]{fontenc}
\usepackage{cite}
\usepackage[cmex10]{amsmath}
\usepackage{amsthm}
\usepackage{amssymb}
\usepackage{bbm}
\usepackage{mathtools}
\usepackage[hidelinks]{hyperref}
\usepackage{algorithmic}
\usepackage{textcomp}
\usepackage[usenames, dvipsnames]{color}
\usepackage[linesnumbered,ruled]{algorithm2e}
\usepackage{subcaption}
\usepackage{graphicx}  
\usepackage{url}
\usepackage[usestackEOL]{stackengine}
\usepackage{verbatim}
\usepackage{dsfont}

\usepackage {tikz}
\tikzset{>=stealth}
\usepackage{xcolor,colortbl}
\definecolor{lightred}{RGB}{255, 240, 240}
\definecolor{lightblue}{RGB}{230, 250, 255}
\definecolor{lightgreen}{RGB}{240, 255, 242}
\definecolor{myred}{RGB}{220, 0, 0}
\definecolor{myblue}{RGB}{0, 17, 173}
\definecolor{mygreen}{RGB}{2, 117, 0}

\newcommand{\mca}[1]{\mathcal{#1}}

\newcommand{\mbf}[1]{\mathbf{#1}}
\newcommand{\mbb}[1]{\mathbb{#1}}
\newcommand{\mrm}[1]{\mathrm{#1}}
\newcommand{\mtt}[1]{\mathtt{#1}}
\newcommand{\bsm}[1]{\boldsymbol{#1}}

\newcommand{\tr}{\operatorname{tr}}

\newcommand{\diag}[1]{\operatorname{diag}{\left(#1\right)}}
\newcommand{\vectorize}[1]{\operatorname{vec}{\left(#1\right)}}

\DeclareMathOperator*{\argmax}{arg\;max}
\DeclareMathOperator*{\maximize}{maximize}

\newcommand{\BS}{\mtt{B}}

\newcommand{\RB}{\mtt{RB}}
\newcommand{\UB}{\mtt{UB}}
\newcommand{\UD}{\mtt{UD}}
\newcommand{\DD}{\mtt{DD}}
\newcommand{\DB}{\mtt{DB}}
\newcommand{\RD}{\mtt{RD}}
\newcommand{\UR}{\mtt{UR}}
\newcommand{\DR}{\mtt{DR}}
\newcommand{\Device}{\mtt{D}}
\newcommand{\User}{\mtt{U}}
\newcommand{\SINR}{\mrm{SINR}}

\makeatletter
\renewcommand*\env@matrix[1][\arraystretch]{%
  \edef\arraystretch{#1}%
  \hskip -\arraycolsep
  \let\@ifnextchar\new@ifnextchar
  \array{*\c@MaxMatrixCols c}}
\makeatother

\ifCLASSINFOpdf
\else
\fi
\usepackage{amsmath}
\begin{document}
%
\title{RIS-Aided Interference Cancellation for Joint Device-to-Device and Cellular Communications}

\author{\IEEEauthorblockN{Ly V. Nguyen and A. Lee Swindlehurst}
\IEEEauthorblockA{Center for Pervasive
Communications and Computing, University of California, Irvine, CA, USA}
Email: vanln1@uci.edu, swindle@uci.edu
\thanks{This work was supported by the National Science Foundation under grants CNS-2107182 and ECCS-2030029.}
}

\maketitle

\begin{abstract}
Joint device-to-device (D2D) and cellular communication is a promising technology for enhancing the spectral efficiency of future wireless networks. However, the interference management problem is challenging since the operating devices and the cellular users share the same spectrum. The emerging reconfigurable intelligent surfaces (RIS) technology is a potentially ideal solution for this interference problem since RISs can shape the wireless channel in desired ways. This paper considers an RIS-aided joint D2D and cellular communication system where the RIS is exploited to cancel interference to the D2D links and maximize the minimum signal-to-interference plus noise (SINR) of the device pairs and cellular users. First, we adopt a popular alternating optimization (AO) approach to solve the minimum SINR maximization problem. Then, we propose an interference cancellation (IC)-based approach whose complexity is much lower than that of the AO algorithm. We derive a representation for the RIS phase shift vector which cancels the interference to the D2D links. Based on this representation, the RIS phase shift optimization problem is transformed into an effective D2D channel optimization. We show that the AO approach can converge faster and can even give better performance when it is initialized by the proposed IC solution. We also show that for the case of a single D2D pair, the proposed IC approach can be implemented with limited feedback from the single receive device.
\end{abstract}



%
\IEEEpeerreviewmaketitle

\section{Introduction}
\label{sec_introduction}
With the proliferation of mobile users and devices, joint device-to-device (D2D) and cellular communication has been envisioned as a promising technology for enhancing spectral efficiency in future wireless networks~\cite{Jameel2018D2D}. In this technology, devices and cellular users operate on the same spectrum, and therefore significantly improve the network spectral efficiency. However, the interference management problem is very challenging due to spectrum sharing between the operating devices and cellular users~\cite{Gu2015Matching,Wang2018Resource,Yang2016Energy}. Fortunately, the D2D interference problem can be effectively addressed by reconfigurable intelligent surfaces (RISs)--a recent innovative technology capable of shaping the wireless channel in desired ways~\cite{Basar2019Wireless,Marco2020Smart}. For example, the works in \cite{Tao2022Interference,Fangzhou2023HRIS,Fangzhou2023ARIS} show that an RIS can be used to effectively null out or mitigate interference in D2D systems. While the work in~\cite{Tao2022Interference} considers a conventional passive RIS structure which can only change the phase of the incoming signal, the authors in~\cite{Fangzhou2023HRIS,Fangzhou2023ARIS} show that a hybrid RIS structure with the capability of simultaneously changing the phase and amplitude of the incoming signal can provide better
interference cancellation performance. However, these prior works all consider systems with only D2D links, i.e., there were no active co-channel cellular users present. 

RIS-assisted joint D2D and cellular communication has been studied in several recent papers. For example, the works in~\cite{Shah2022Statistical} and~\cite{Yiyang2021Performance} perform analyses for RIS-assisted D2D communications under Rician and Nakagami-$m$ fading channels, respectively. The study in~\cite{Shah2022Statistical} investigates the effective capacity and performs mode selection to determine whether the D2D communication is established through an RIS or a base station (BS). The authors of~\cite{Yiyang2021Performance} derived closed-form expressions for the outage probability of the D2D links and the resulting diversity order. Unlike~\cite{Shah2022Statistical,Yiyang2021Performance} which perform system analysis, some other recent papers focus on system design and optimization~\cite{Zelin2022RIS,Fangmin2022Sumrate,sultana2022intelligent,Yali2021RIS,Yashuai2021Sumrate,Gang2021RIS}. In particular, the sum-rate maximization problem was considered in~\cite{Zelin2022RIS,Fangmin2022Sumrate,sultana2022intelligent,Yali2021RIS} and deep reinforcement learning was used in~\cite{Zelin2022RIS,sultana2022intelligent}. However, the works in~\cite{Zelin2022RIS,sultana2022intelligent} assume that the BS is equipped with only one antenna and only one D2D pair is allowed to share the spectrum with a single cellular user. Multiple D2D pairs sharing the same spectrum with a cellular user were considered in~\cite{Fangmin2022Sumrate,Yali2021RIS}, but it was still assumed that the BS has only one antenna. The sum-rate maximization problem for systems with a multi-antenna BS was studied in~\cite{Yashuai2021Sumrate,Gang2021RIS}, but again they assume only one D2D pair sharing the spectrum with one cellular user.

Unlike the aforementioned works which consider the sum-rate maximization problem with relaxed system assumptions, we study a different problem that takes into account the fairness by maximizing the minimum signal-to-interference-plus-noise ratio (SINR) over the cellular users and devices. We also consider a more challenging RIS-assisted joint D2D and cellular communication system where multiple cellular users share the same spectrum with multiple D2D pairs and the BS is also equipped with multiple antennas. The contributions of our paper is summarized as follows:
\begin{itemize}
    \item First, we adopt a popular alternating optimization (AO) approach and the semi-definite relaxation (SDR) technique to optimize the combining matrix at the BS and the RIS phase shift vector that maximize the minimum SINR over the cellular users and devices.
    \item Next, we propose an interference cancellation (IC)-based approach whose complexity is much lower than that of the AO algorithm. More specifically, we first derive a representation for the RIS phase shift vector that cancels the interference to the D2D links. Based on this representation, we transform the RIS phase shift optimization problem into an effective D2D channel optimization problem. We show that the AO approach can converge faster and can even give better performance when it uses the proposed IC solution as its initial point.
    \item We also show that for the case of a single D2D pair, the proposed IC approach can be implemented with limited feedback from the receive device. Finally, we numerically show the gains and benefits of the proposed approaches.
\end{itemize}


\section{System Model and Problem Formulation}
\label{sec_system_model}
\begin{table}[t!]
\centering
\renewcommand{\arraystretch}{1.4}
\caption{Channel description and notation.}
\label{table:channel_notation}
\begin{tabular}{||l|c|l||}
\hline
 \textbf{Channel} & \textbf{Notation} &  \textbf{Size}\\ \hline
RIS $\to$ BS & $\mbf{H}^{\RB}$  &  $M \times N$ \\ 
CU $\to$ RxD & $\mbf{H}^{\UD}$ & $L \times K$ \\
TxD $\to$ RxD & $\mbf{H}^{\DD}$ & $L \times L$ \\
CU $\to$ BS & $\mbf{H}^{\UB} \overset{\Delta}{=} [\mbf{h}^{\UB}_1,\ldots,\mbf{h}^{\UB}_K]$ & $M \times K$ \\
TxD $\to$ BS & $\mbf{H}^{\DB} \overset{\Delta}{=} [\mbf{h}^{\DB}_1,\ldots,\mbf{h}^{\DB}_L]$ &  $M \times L$ \\
RIS $\to$ RxD& $\mbf{H}^{\RD} \overset{\Delta}{=} [\mbf{h}^{\RD}_1,\ldots,\mbf{h}^{\RD}_L]^H$ & $L \times N$ \\  
CU $\to$ RIS & $\mbf{H}^{\UR} \overset{\Delta}{=} [\mbf{h}^{\UR}_1,\ldots,\mbf{h}^{\UR}_K]$ & $N \times K$ \\ 
TxD $\to$ RIS & $\mbf{H}^{\DR} \overset{\Delta}{=} [\mbf{h}^{\DR}_1,\ldots,\mbf{h}^{\DR}_L]$ & $N \times L$ \\ 
 \hline
cascaded TxD$\to$RxD & $(\mbf{g}_{\ell,\ell'}^{\DD})^H = (\mbf{h}_{\ell}^{\RD})^H\diag{\mbf{h}_{\ell'}^{\DR}}$ & $1 \times N$ \\ \hline
cascaded CU$\to$RxD& $(\mbf{g}_{\ell,k}^{\UD})^H = (\mbf{h}_{\ell}^{\RD})^H\diag{\mbf{h}_{k}^{\UR}}$ & $1 \times N$ \\ \hline
cascaded CU$\to$BS& $\mbf{G}_{k}^{\UB} = \mbf{H}^{\RB}\diag{\mbf{h}_{k}^{\UR}}$ & $M \times N$ \\ \hline
cascaded TxD$\to$BS& $\mbf{G}_{\ell}^{\DB} = \mbf{H}^{\RB}\diag{\mbf{h}_{\ell}^{\DR}}$ & $M \times N$ \\
\hline
\end{tabular}
\end{table}
\subsection{System Model}
We consider an uplink RIS-assisted MIMO system where an $M$-antenna base station serves $K$ users with the assistance of an $N$-element RIS. There are also $L$ device pairs sharing the same spectrum with the cellular system. We assume that the users and devices are all equipped with a single antenna and the transmit power of user-$k$ and device-$\ell$ are denoted as $p^\User_k$ and $p^\Device_\ell$, respectively. Since the BS has multiple antennas serving multiple users, a combining matrix $\mbf{W} = [\mbf{w}_1, \ldots, \mbf{w}_K]^H$ is applied to the BS's received signal to recover the cellular users' data. Here, $\mbf{w}_k$ is the combiner for user-$k$. The receive devices have only one antenna, so they directly detect data from their received signal. The RIS is defined by a vector $\bsm{\phi} = [\phi_1, \ldots,\phi_N]^T$ where $|\phi_i| \leq 1$. Table~\ref{table:channel_notation} provides a description of all channels and their corresponding mathematical definition. In this paper, we assume perfect channel state estimation information (CSI) at the BS and the receive devices. However, we do not assume knowledge of the individual channels to and from the RIS but their cascaded versions, the estimation of which has been intensively studied in the RIS literature. The receive devices feedback their CSI to the BS for optimizing the combining matrix $\mbf{W}$ and the RIS phase shift vector $\bsm{\phi}$. It is also assumed that the noise at the BS and at the receive devices is distributed as $\mca{CN}(\mbf{0}_M, \sigma^2_\mtt{B}\mbf{I}_M)$ and $\mca{CN}(0,\sigma^2_\mtt{D})$, respectively.

\subsection{Problem Formulation}
The problem of interest is to maximize the minimum SINR:
\begin{equation}
\begin{aligned}
& \maximize_{\{\bsm{\phi},\, \mbf{w}_k\}}
& & \min_{k,\ell} \; \{\SINR_\ell^\Device,\, \SINR_k^\User\}\\
& \operatorname{subject\ to}
&& |\phi_i| \leq 1, \; \forall i = 1,\ldots, N.
\end{aligned}
\label{eq:problem}
\end{equation}
where $\SINR_\ell^\Device$ and $\SINR_k^\User$ are given in~\eqref{eq:SINR_Device} and~\eqref{eq:SINR_User} and denote the SINR of the $\ell$-th device pair and the SINR of user-$k$, respectively.
	\begin{figure*}
   \begin{equation}
      \SINR_{\ell}^{\Device} =  \frac{p_{\ell}^\Device|h_{\ell,\ell}^{\DD} + (\mbf{g}_{\ell,\ell}^\DD)^H\bsm{\phi}|^2}{\sum_{\ell'\neq \ell} p_{\ell'}^\Device|h_{\ell,\ell'}^{\DD} + (\mbf{g}_{\ell,\ell'}^\DD)^H\bsm{\phi}|^2+\sum_{k=1}^K p_{k}^\User|h_{\ell,k}^{\UD} + (\mbf{g}_{\ell,k}^\UD)^H\bsm{\phi}|^2 + \sigma^2_{\Device}}
      \label{eq:SINR_Device}
  \end{equation}
		\begin{eqnarray}
  \SINR_{k}^{\User} =  \frac{p_{k}^\User|\mbf{w}_k^H(\mbf{h}_{k}^{\UB} + \mbf{G}_{k}^\UB\bsm{\phi})|^2}{\sum_{k'\neq k} p_{k'}^\User|\mbf{w}_k^H(\mbf{h}_{k'}^{\UB} + \mbf{G}_{k'}^\UB\bsm{\phi})|^2+\sum_{\ell=1}^L p_{\ell}^\Device|\mbf{w}_k^H(\mbf{h}_{\ell}^{\DB} + \mbf{G}_{\ell}^\DB\bsm{\phi})|^2 + \|\mbf{w}_k\|^2\sigma^2_{\BS}}
  \label{eq:SINR_User}
		\end{eqnarray}
		\hrulefill
	\end{figure*}

\section{Alternating Optimization Approach}
This section develops an AO approach to solve the joint D2D and cellular communication problem in~\eqref{eq:problem}. Here, we alternately optimize the BS combiners $\{\mbf{w}_k\}$ and the RIS phase shift $\bsm{\phi}$.

\subsection{Optimizing $\{\mbf{w}_k\}$ given $\bsm{\phi}$}
Let $\mbf{f}_k^\UB \overset{\Delta}{=} \mbf{h}_{k}^{\UB} + \mbf{G}_{k}^\UB\bsm{\phi}$ and $\mbf{f}_\ell^\DB \overset{\Delta}{=} \mbf{h}_{\ell}^{\DB} + \mbf{G}_{\ell}^\DB\bsm{\phi}$ define the effective channel from user-$k$ to the BS and from the transmit device-$\ell$ to the BS, respectively. The optimal linear minimum mean squared error (LMMSE) combiner at the BS for user-$k$ is given as 
\begin{equation*}
    \mbf{w}_k = \bigg(\sum_{k'=1}^Kp^\User_{k'}\mbf{f}_{k'}^\UB(\mbf{f}_{k'}^\UB)^H + \sum_{\ell=1}^Lp^\Device_{\ell}\mbf{f}_\ell^\DB(\mbf{f}_\ell^\DB)^H+\sigma^2_\BS\mbf{I}_M\bigg)^{-1}\mbf{f}^\UB_k.
\end{equation*}

\subsection{Optimizing $\bsm{\phi}$ given $\{\mbf{w}_k\}$}
Let
$$
(\bsm{\psi}^\UB_{k,k'})^H \overset{\Delta}{=} \mbf{w}_k^H\mbf{G}_{k'}^\UB\;\;\text{and}\;\;  \alpha^\UB_{k,k'} \overset{\Delta}{=} \mbf{w}_k^H\mbf{h}_{k'}^{\UB},
$$
then we can express the term $|\mbf{w}_k^H(\mbf{h}_{k'}^{\UB} + \mbf{G}_{k'}^\UB\bsm{\phi})|^2$ in~\eqref{eq:SINR_User} as follows:
\begin{align}
    |\mbf{w}_k^H(\mbf{h}_{k'}^{\UB} + \mbf{G}_{k'}^\UB\bsm{\phi})|^2 = 
    \bsm{\tilde{\phi}}^H \bsm{\Psi}^\UB_{k,k'} \bsm{\tilde{\phi}} = \tr(\bsm{\tilde{\Phi}}\bsm{\Psi}^\UB_{k,k'})
    \label{eq:UB_power}
\end{align}
where $\bsm{\Tilde{\phi}} = [\bsm{\phi}^T, 1]^T$, $\bsm{\Tilde{\Phi}} = \bsm{\Tilde{\phi}}\bsm{\Tilde{\phi}}^H$, and
\begin{equation*}
    \bsm{\Psi}^\UB_{k,k'} = \begin{bmatrix}[1.2]
    \bsm{\psi}^\UB_{k,k'}(\bsm{\psi}^\UB_{k,k'})^H & \alpha^\UB_{k,k'}\bsm{\psi}^\UB_{k,k'} \\
    (\bsm{\psi}^\UB_{k,k'})^H(\alpha^\UB_{k,k'})^* & |\alpha^\UB_{k,k'}|^2
\end{bmatrix}.
\end{equation*}

Similarly, by defining $$(\bsm{\psi}^\DB_{k,\ell})^H \overset{\Delta}{=} \mbf{w}_k^H\mbf{G}_{\ell}^\DB\;\; \text{and}\;\; \alpha^\DB_{k,\ell} \overset{\Delta}{=} \mbf{w}_k^H\mbf{h}_{\ell}^{\DB},$$
we can write the term $|\mbf{w}_k^H(\mbf{h}_{\ell}^{\DB} + \mbf{G}_{\ell}^\DB\bsm{\phi})|^2$ in~\eqref{eq:SINR_User} as
\begin{align}
    |\mbf{w}_k^H(\mbf{h}_{\ell}^{\DB} + \mbf{G}_{\ell}^\DB\bsm{\phi})|^2 &= \bsm{\tilde{\phi}}^H \bsm{\Psi}^\DB_{k,\ell} \bsm{\tilde{\phi}} = \tr(\bsm{\tilde{\Phi}}\bsm{\Psi}^\DB_{k,\ell})
    \label{eq:DB_power}
\end{align}
where
\begin{equation*}
    \bsm{\Psi}^\DB_{k,\ell} = \begin{bmatrix}[1.2]
    \bsm{\psi}^\DB_{k,\ell}(\bsm{\psi}^\DB_{k,\ell})^H & \alpha^\DB_{k,\ell}\bsm{\psi}^\DB_{k,\ell} \\
    (\bsm{\psi}^\DB_{k,\ell})^H(\alpha^\DB_{k,\ell})^* & |\alpha^\DB_{k,\ell}|^2
\end{bmatrix}.
\end{equation*}

Using~\eqref{eq:UB_power} and~\eqref{eq:DB_power}, the SINR of user-$k$ in~\eqref{eq:SINR_User} is written as
\begin{equation}
 \label{eq:SINR_Uk}
    \SINR_{k}^{\User}  = \frac{p_{k}^\User \tr(\bsm{\tilde{\Phi}}\bsm{\Psi}^\UB_{k,k})}{\sum_{k'\neq k}p_{k'}^\User \tr(\bsm{\tilde{\Phi}}\bsm{\Psi}^\UB_{k,k'}) + \sum_{\ell} p_{\ell}^\Device \tr(\bsm{\tilde{\Phi}}\bsm{\Psi}^\DB_{k,\ell}) + \zeta_k}
\end{equation}
where $\zeta_k = \|\mbf{w}_k\|^2\sigma^2_{\BS}$.
We also have
 \begin{equation}
    \label{eq:SINR_Dl}
      \SINR_{\ell}^{\Device} =  \frac{p_{\ell}^\Device\tr(\bsm{\tilde{\Phi}}\bsm{\Psi}^\DD_{\ell,\ell})}{\sum_{\ell'\neq \ell} p_{\ell'}^\Device\tr(\bsm{\tilde{\Phi}}\bsm{\Psi}^\DD_{\ell,\ell'})+\sum_{k} p_{k}^\User\tr(\bsm{\tilde{\Phi}}\bsm{\Psi}^\UD_{\ell,k}) + \sigma^2_{\Device}}
  \end{equation}
where \begin{align*}
    \bsm{\Psi}^\DD_{\ell,\ell'} &= \begin{bmatrix}[1.2]
    \mbf{g}^\DD_{\ell,\ell'}(\mbf{g}^\DD_{\ell,\ell'})^H & h^\DD_{\ell,\ell'}\mbf{g}^\DD_{\ell,\ell'} \\
    (\mbf{g}^\DD_{\ell,\ell'})^H(h^\DD_{\ell,\ell'})^* & |h^\DD_{\ell,\ell'}|^2
\end{bmatrix}, \\
\bsm{\Psi}^\UD_{\ell,k} &= \begin{bmatrix}[1.2]
    \mbf{g}^\UD_{\ell,k}(\mbf{g}^\UD_{\ell,k})^H & h^\UD_{\ell,k}\mbf{g}^\UD_{\ell,k} \\
    (\mbf{g}^\UD_{\ell,k})^H(h^\UD_{\ell,k})^* & |h^\UD_{\ell,k}|^2
\end{bmatrix}.
\end{align*}

Therefore, problem~\eqref{eq:problem} can be written as
\begin{equation}
\begin{aligned}
& \maximize_{\{\xi, \; \bsm{\Tilde{\Phi}} \succeq \mbf{0} \}}
& & \xi\\
& \operatorname{subject\ to}
&& \text{\eqref{eq:SINR_Uk}} \geq \xi \; \forall k\\
&&& \text{\eqref{eq:SINR_Dl}} \geq \xi \;  \forall \ell \\
&&& |\bsm{\tilde{\Phi}}_{i,i}| \leq 1, \; \forall i = 1, \ldots, N \\
&&& \bsm{\tilde{\Phi}}_{N+1,N+1} = 1.
\end{aligned}
\label{eq:AO_phi_optimize}
\end{equation}
This is a fractional programming problem and thus can be solved by the Dinkelbach algorithm~\cite{Fangshou2020Joint}. Given a solution $\bsm{\hat{\Tilde{\Phi}}}$ of~\eqref{eq:AO_phi_optimize}, we need to convert $\bsm{\hat{\Tilde{\Phi}}}$ into a feasible solution $\bsm{\hat{\phi}}$, which can be done by the Gaussian randomization approach~\cite{Luo2010SDR}. Specifically, we first generate a set of Gaussian random vectors as $\bsm{{\Tilde{\phi}}}_i \sim \mca{CN}(\mbf{0},\bsm{\hat{\Tilde{\Phi}}})$, $i = 1, \ldots, I$. Since the elements of $\bsm{{\Tilde{\phi}}}_i$ may not satisfy the constraints $|\tilde{\phi}_{i,n}| \leq 1$ for $n = 1, \ldots, N$ and $\tilde{\phi}_{i,N+1} = 1$, we need to first normalize $\bsm{{\Tilde{\phi}}}_i$ to obtain $\bsm{{\bar{\phi}}}_i = \bsm{{\Tilde{\phi}}}_i/\tilde{\phi}_{i,N+1}$, then we can obtain a feasible candidate $\bsm{\hat{\phi}}_{i} = [\hat{\phi}_{i,1},\ldots,\hat{\phi}_{i,N}]^T$ where $\hat{\phi}_{i,n} = e^{j\measuredangle(\bar{\phi}_{i,n})}$ if $|\bar{\phi}_{i,n}| > 1$, otherwise $\hat{\phi}_{i,n} = \tilde{\phi}_{i,n}$. Finally, a solution $\bsm{\hat{\phi}}$ of problem~\eqref{eq:AO_phi_optimize} can be obtained as
\begin{equation}
    \bsm{\hat{\phi}} = \argmax_{\{\bsm{\hat{\phi}}_{1},\ldots,\bsm{\hat{\phi}}_{I}\}} \;\;\min \, \{\SINR_\ell^\Device,\, \SINR_k^\User\}.
    \label{eq:Gauss_rand_solution}
\end{equation}

\section{Interference Cancellation Approach}
The AO approach presented above is efficient but its computational complexity is high because the BS combiners and the RIS phase shift vector are alternately optimized. In addition, each AO iteration requires use of the Dinkelbach algorithm which is an iterative algorithm itself, and each of its iterations also requires the solution of an SDR optimization problem. Here, we propose an IC approach whose computational complexity is significantly lower than that of the AO algorithm since the proposed IC approach only needs to solve one SDR problem. Details of the proposed IC approach are presented as follows.

Let $\mbf{F}^\DD \in \mbb{C}^{L\times L}$ be the effective channel from the transmit devices to the receive devices, and define
\begin{align}
    \mbf{f}^\DD = \vectorize{(\mbf{F}^\DD)^T} 
    &= \vectorize{(\mbf{H}^\DD)^T} + \begin{bmatrix}[0.8]
        \mbf{G}^\DD_1\\ \vdots \\
        \mbf{G}^\DD_L
    \end{bmatrix}\bsm{\phi} \notag\\
    &= \mbf{h}^\DD + \mbf{G}^\DD\bsm{\phi},
\end{align}
where $\mbf{G}^\DD_\ell = [\mbf{g}^\DD_{\ell,1},\ldots,\mbf{g}^\DD_{\ell,L}]^H$. We use a permutation matrix $\mbf{P}$ to separate the vector $\mbf{f}^\DD$ into two sub-vectors $\mbf{f}^\DD_{\mrm{sig}}$ and $\mbf{f}^\DD_{\mrm{itf}}$ which contain the diagonal and off-diagonal elements of $\mbf{F}^\DD$, respectively. The diagonal elements of $\mbf{F}^\DD$ represent the useful channels for the D2D pairs, while the off-diagonal elements represent the interference channels from other D2D pairs. Hence we use the subscripts `$\mrm{sig}$' and `$\mrm{itf}$' to indicate the D2D signal and interference, respectively. This means we can write
\begin{equation}
    \begin{bmatrix}[1.2]
        \mbf{f}^\DD_{\mrm{sig}} \\
        \mbf{f}^\DD_{\mrm{itf}}
    \end{bmatrix} = \begin{bmatrix}[1.2]
        \mbf{h}^\DD_{\mrm{sig}} \\
        \mbf{h}^\DD_{\mrm{itf}}
    \end{bmatrix} + \begin{bmatrix}[1.2]
        \mbf{G}^\DD_{\mrm{sig}}\\
        \mbf{G}^\DD_{\mrm{itf}}
    \end{bmatrix} \bsm{\phi}
    \label{eq:f_DD_decompose}
\end{equation}
where
\begin{equation*}
    \begin{bmatrix}[1.2]
        \mbf{f}^\DD_{\mrm{sig}} \\
        \mbf{f}^\DD_{\mrm{itf}}
    \end{bmatrix} = \mbf{P}\mbf{f}^\DD, \;\; \begin{bmatrix}[1.2]
        \mbf{h}^\DD_{\mrm{sig}} \\
        \mbf{h}^\DD_{\mrm{itf}}
    \end{bmatrix} =\mbf{P}\mbf{h}^\DD,\;\;  \begin{bmatrix}[1.2]
        \mbf{G}^\DD_{\mrm{sig}}\\
        \mbf{G}^\DD_{\mrm{itf}}
    \end{bmatrix} =\mbf{P}\mbf{G}^\DD.
\end{equation*}
Similarly, let $\mbf{F}^\UD \in \mbb{C}^{L\times K}$ be the effective channel from the cellular users to the receive devices, so we can also write
\begin{align}
    \mbf{f}^\UD = \vectorize{(\mbf{F}^\UD)^T} 
    &= \vectorize{(\mbf{H}^\UD)^T} + \begin{bmatrix}[0.8]
        \mbf{G}_1^\UD\\ \vdots \\
        \mbf{G}_L^\UD
    \end{bmatrix}\bsm{\phi} \notag\\
    &= \mbf{h}^\UD + \mbf{G}^\UD\bsm{\phi},
    \label{eq:f_UD}
\end{align}
where $\mbf{G}_\ell^\UD = [\mbf{g}_{\ell,1}^{\UD},\ldots,\mbf{g}_{\ell,K}^{\UD}]^H$.

Combining~\eqref{eq:f_DD_decompose} and~\eqref{eq:f_UD}, we have
\begin{equation}
    \begin{bmatrix}[1.2]
        \mbf{f}^\DD_{\mrm{sig}} \\
        \mbf{f}^\DD_{\mrm{itf}} \\
        \mbf{f}^\UD
    \end{bmatrix} = \begin{bmatrix}[1.2]
        \mbf{h}^\DD_{\mrm{sig}} \\
        \mbf{h}^\DD_{\mrm{itf}} \\
        \mbf{h}^\UD
    \end{bmatrix} + \underbrace{\begin{bmatrix}[1.2]
        \mbf{G}^\DD_{\mrm{sig}}\\
        \mbf{G}^\DD_{\mrm{itf}} \\
        \mbf{G}^\UD
    \end{bmatrix}}_{\mbf{A}} \bsm{\phi},
    \label{eq:f_D}
\end{equation}
where $\mbf{A} \in \mbb{C}^{L(K+L) \times N}$ is a matrix representing the RIS-aided channel. Assuming that $N \geq L(K+L)$, then using~\eqref{eq:f_D} we can write the phase shift vector $\bsm{\phi}$ in the following expression: 
\begin{equation}
    \bsm{\phi} = \mbf{A}^H(\mbf{A}\mbf{A}^H)^{-1}\left(\begin{bmatrix}[1.2]
        \mbf{f}^\DD_{\mrm{sig}} \\
        \mbf{f}^\DD_{\mrm{itf}} \\
        \mbf{f}^\UD
    \end{bmatrix} - \begin{bmatrix}[1.2]
        \mbf{h}^\DD_{\mrm{sig}} \\
        \mbf{h}^\DD_{\mrm{itf}} \\
        \mbf{h}^\UD
    \end{bmatrix}\right).
    \label{eq:phi}
\end{equation} 
    
Let $[\mbf{B}, \mbf{C}] \overset{\Delta}{=}\mbf{A}^H(\mbf{A}\mbf{A}^H)^{-1}$ where $\mbf{B}\in\mbb{C}^{N\times L}$ and $\mbf{C}\in\mbb{C}^{N \times L(K+L)-L}$, then decompose the right-hand side of~\eqref{eq:phi} as follows:
\begin{align}
    \bsm{\phi} &= \mbf{B}\mbf{f}^\DD_{\mrm{sig}} - \mbf{B}\mbf{h}^\DD_{\mrm{sig}} + \mbf{C}\left(\begin{bmatrix}[1.2]
        \mbf{f}^\DD_{\mrm{itf}} \\
        \mbf{f}^\UD
    \end{bmatrix} - \begin{bmatrix}[1.2]
        \mbf{h}^\DD_{\mrm{itf}} \\
        \mbf{h}^\UD
    \end{bmatrix}\right).
    \label{eq:phi_decompose}
\end{align}
The representation above shows a relationship between the RIS phase shift $\bsm{\phi}$ and the effective device-related channels including the useful D2D channels $\mbf{f}^\DD_{\mrm{sig}}$ and the other interference links $\mbf{f}^\DD_{\mrm{itf}}$ and $\mbf{f}^\UD$. If we let $\mbf{f}^\DD_{\mrm{itf}} = \mbf{0}$ and $\mbf{f}^\UD = \mbf{0}$, the RIS solution~\eqref{eq:phi_decompose} reduces to an IC $\bsm{\phi}$ solution expressed in terms of $\mbf{f}^\DD_{\mrm{sig}}$ as
\begin{align}
    \bsm{\phi}_{\mrm{IC}} &= \mbf{B}\mbf{f}^\DD_{\mrm{sig}} - \mbf{B}\mbf{h}^\DD_{\mrm{sig}} - \mbf{C}\begin{bmatrix}[1.2]
        \mbf{h}^\DD_{\mrm{itf}} \\
        \mbf{h}^\UD
    \end{bmatrix}\notag \\
    &= \mbf{B}\mbf{f}^\DD_{\mrm{sig}} + \mbf{d} \label{eq:phi_IC}
\end{align}
where $\mbf{d} = - \mbf{B}\mbf{h}^\DD_{\mrm{sig}} - \mbf{C}[(\mbf{h}^\DD_{\mrm{itf}})^T, \, (\mbf{h}^\UD)^T]^T$. This means that if we could design the RIS phase shifts with~\eqref{eq:phi_IC}, the interference to the receive devices will be completely cancelled.  In this case, the SINR at the receive device-$\ell$ would simplify to $\SINR_\ell^\Device = p^\Device_\ell|f_{\mrm{sig},\ell}^\DD|^2/\sigma_\Device^2$.

With the $\bsm{\phi}_{\mrm{IC}}$ representation in~\eqref{eq:phi_IC}, we can use the Cauchy–Schwarz inequality to find the following upper bound for the SINR of user-$k$:
\begin{align}
    \SINR_{k}^{\User}  &\leq p^\User_k\|\mbf{h}^\UB_k + \mbf{G}^\UB_k \bsm{\phi}_{\mrm{IC}}\|^2/\sigma_\BS^2 \\
    &= p^\User_k\|\mbf{h}^\UB_k + \mbf{G}^\UB_k (\mbf{B}\mbf{f}^\DD_{\mrm{sig}} + \mbf{d})\|^2/\sigma_\BS^2\\
    &= p^\User_k\|\mbf{Z}_k\mbf{f}^\DD_{\mrm{sig}} + \mbf{q}_k\|^2/\sigma_\BS^2,
\end{align}
where $\mbf{Z}_k = \mbf{G}^\UB_k \mbf{B}$ and $\mbf{q}_k = \mbf{h}^\UB_k + \mbf{G}^\UB_k\mbf{d}$. Therefore, we propose the following IC-based optimization problem
\begin{equation}
\begin{aligned}
& \maximize_{\{\xi, \; \mbf{f}_{\mrm{sig}}^\DD\}}
& & \xi\\
& \operatorname{subject\ to}
&& p^\User_k\|\mbf{Z}_k\mbf{f}^\DD_{\mrm{sig}} + \mbf{q}_k\|^2/\sigma_\BS^2 \geq \xi \; \forall k\\
&&& p^\Device_\ell|f_{\mrm{sig},\ell}^\DD|^2/\sigma_\Device^2 \geq \xi \; \forall \ell \\
&&& |\mbf{b}_i^H\mbf{f}_{\mrm{sig}}^\DD + d_i|^2 \leq 1 \; \forall i = 1,\ldots, N,
\end{aligned}
\label{eq:transformed_prob}
\end{equation}
where the constraint $|\mbf{b}_i^H\mbf{f}_{\mrm{sig}}^\DD + d_i|^2 \leq 1$ is equivalent to the constraint $|\phi_i|^2 \leq 1$. Hence, we have transformed the RIS phase shift $\bsm{\phi}$ optimization problem into an effective D2D channel optimization~\eqref{eq:transformed_prob} by exploiting the IC representation~\eqref{eq:phi_IC}.

The problem in~\eqref{eq:transformed_prob} can also be easily solved using SDR. Specifically, let
\begin{equation*}
    \bsm{\Omega}_{k} = \begin{bmatrix}[1.2]
    \mbf{Z}_{k}^H\mbf{Z}_{k} & \mbf{Z}_{k}^H\mbf{q}_{k} \\
    \mbf{q}_{k}^H\mbf{Z}_{k} & \|\mbf{q}_{k}\|^2
\end{bmatrix} \;\text{and}\; \bsm{\Upsilon}_{i} = \begin{bmatrix}[1.2]
    \mbf{b}_{i}\mbf{b}_{i}^H & \mbf{b}_{i}d_i \\
    \mbf{b}_{i}^Hd_i^* & |d_i|^2
\end{bmatrix} , 
\end{equation*}
so that
\begin{align*}
    &\|\mbf{Z}_k\mbf{f}^\DD_{\mrm{sig}} + \mbf{q}_k\|^2 = \tr(\mbf{\tilde{F}}^\DD_{\mrm{sig}}\bsm{\Omega}_k)\\
    &|\mbf{b}_i^H\mbf{h}_{\mrm{sig}}^\DD + d_i|^2 = \tr(\mbf{\tilde{F}}^\DD_{\mrm{sig}}\bsm{\Upsilon}_i),
\end{align*}
where $\mbf{\tilde{F}}^\DD_{\mrm{sig}} = \mbf{\tilde{f}}^\DD_{\mrm{sig}} (\mbf{\tilde{f}}^\DD_{\mrm{sig}})^T$ with $\mbf{\tilde{f}}^\DD_{\mrm{sig}} = [(\mbf{{f}}^\DD_{\mrm{sig}})^T,1]^T$. Problem~\eqref{eq:transformed_prob} can be now relaxed to a convex optimization problem:
\begin{equation}
\begin{aligned}
& \maximize_{\{\xi, \; \mbf{F}_{\mrm{sig}}^\DD \succeq \mbf{0}\}}
& & \xi\\
& \operatorname{subject\ to}
&& p^\User_k\tr(\mbf{\tilde{F}}^\DD_{\mrm{sig}}\bsm{\Omega}_k)/\sigma_\BS^2 \geq \xi \; \forall k\\
&&& p^\Device_\ell\Re\{\mbf{\tilde{F}}^\DD_{\mrm{sig},\ell,\ell}\}/\sigma_\Device^2 \geq \xi \; \forall \ell \\
&&& \tr(\mbf{\tilde{F}}^\DD_{\mrm{sig}}\bsm{\Upsilon}_i) \leq 1 \; \forall i = 1,\ldots, N \\
&&& \mbf{\tilde{F}}^\DD_{\mrm{sig},L+1,L+1} = 1,
\end{aligned}
\end{equation}
which again can be solved using SDR. Finally, we need to recover an IC solution $\bsm{\hat{\phi}}_{\mrm{IC}}$ from $\mbf{\tilde{F}}^\DD_{\mrm{sig}}$, which can also be done using the Gaussian randomization method. Given a $\bsm{\hat{\phi}}_{\mrm{IC}}$, the corresponding LMMSE combiner at the BS can be computed. 

It can be seen that unlike the AO approach which needs to solve a series of SDR problems, the IC approach involves only one SDR optimization and therefore significantly reduces the computational complexity. We will show later that if the AO approach is initialized with $\bsm{\hat{\phi}}_{\mrm{IC}}$, it converges more rapidly with better performance compared to using a random initialization.

\textbf{Limited feedback for single D2D pair:}
To implement the IC method for the case of multiple device pairs, it is required that each receive device-$\ell$ forwards its estimated CSI including two cascaded channel matrices $\mbf{G}_\ell^\DD \in \mbb{C}^{L\times N}$, $\mbf{G}_\ell^\UD \in \mbb{C}^{K\times N}$ and two direct channel vectors $\mbf{H}^\DD_{\ell,:} \in \mbb{C}^{1\times L}$, $\mbf{H}^\UD_{\ell,:} \in \mbb{C}^{1\times K}$ via a control link to the BS, which results in a total number of $(L+K)(N+1)$ complex coefficients to be fed-back.

However, for the case of a single device pair, the representation in~\eqref{eq:phi} can be obtained at the receive device as follows:
\begin{align}
    \bsm{{\phi}} &= \mbf{A}^H(\mbf{A}\mbf{A}^H)^{-1}\left(\begin{bmatrix}[1.2]
        f^\DD \\
        \mbf{f}^\UD
        \end{bmatrix} - \begin{bmatrix}[1.2]
        h^\DD \\
        \mbf{h}^\UD
    \end{bmatrix}\right)
    \label{eq:phi_decompose_1D}
\end{align}
where $\mbf{A} = [(\mbf{g}^\DD)^T,\,\mbf{G}^\UD]$ and we have dropped the device index $\ell$ since there is only D2D pair here. By setting $\mbf{f}^\UD = \mbf{0}$, the $\bsm{\phi}_{\mrm{IC}}$ representation that cancels the interference at the device is
\begin{align}
    \bsm{{\phi}}_{\mrm{IC}} &= \mbf{b} f^\DD + \mbf{d},
\end{align}
where $\mbf{b}$ is the first column of $\mbf{A}^H(\mbf{A}\mbf{A}^H)^{-1}$ and $\mbf{d} = -\mbf{A}^H(\mbf{A}\mbf{A}^H)^{-1}[h^\DD, (\mbf{h}^\UD)^T]^T$. This means the receive device can calculate the matrix $\mbf{A}^H(\mbf{A}\mbf{A}^H)^{-1}$ using its local CSI, then obtain the two corresponding vectors $\mbf{b}$ and $\mbf{d}$, which will be forwarded to the BS to solve the RIS phase shift optimization problem:
\begin{equation}
\begin{aligned}
& \maximize_{\{\xi, \; {f}^\DD\}}
& & \xi\\
& \operatorname{subject\ to}
&& p^\User_k\|\mbf{z}_k{f}^\DD + \mbf{q}_k\|^2/\sigma_\BS^2 \geq \xi \; \forall k\\
&&& p^\Device|f^\DD|^2/\sigma_\Device^2 \geq \xi \\
&&& |b_if^\DD + d_{i}|^2 \leq 1 \; \forall i = 1,\ldots, N.
\end{aligned}
\end{equation}

Thus, instead of forwarding the $(K+1)(N+1)$ complex coefficients from $\mbf{g}^\DD \in \mbb{C}^{N\times 1}$, $\mbf{G}^\UD \in \mbb{C}^{K\times N}$, $h^\DD$, and $\mbf{h}^\UD \in \mbb{C}^{K\times 1}$, the receive device finds $\mbf{b}$ and $\mbf{d}$ and forwards these two vectors to the BS. This only requires transmission of $2N$ complex coefficients, but still guarantees that the BS can solve the RIS phase shift optimization problem.

\begin{figure}[t!]
    \centering
    \begin{subfigure}[t]{\linewidth}
        \centering
        \includegraphics[width=0.95\linewidth]{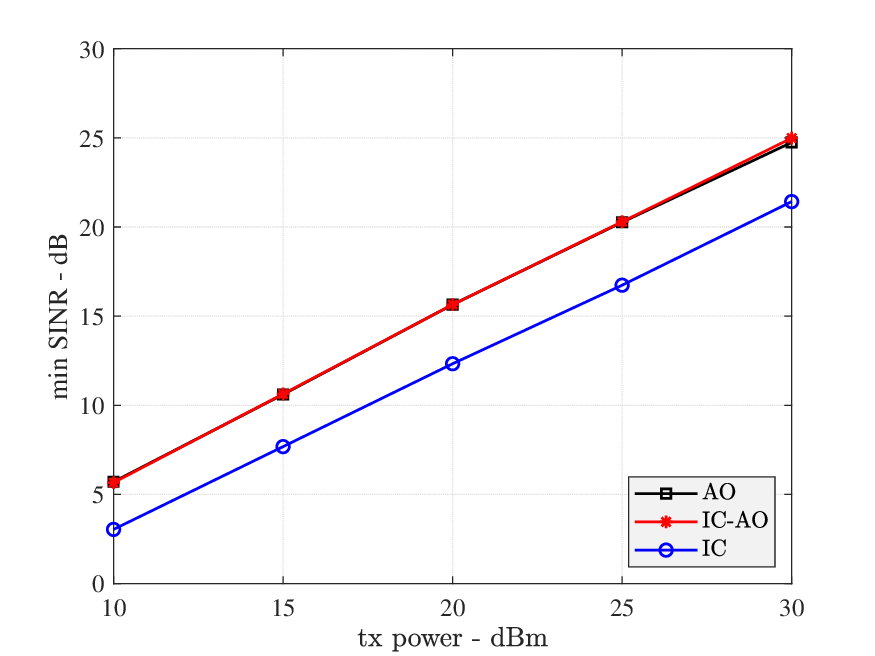}
        \caption{Minimum SINR performance}
        \label{fig_min_SINR_vs_P}
    \end{subfigure}
    
    \begin{subfigure}[t]{\linewidth}
        \centering
        \includegraphics[width=0.95\linewidth]{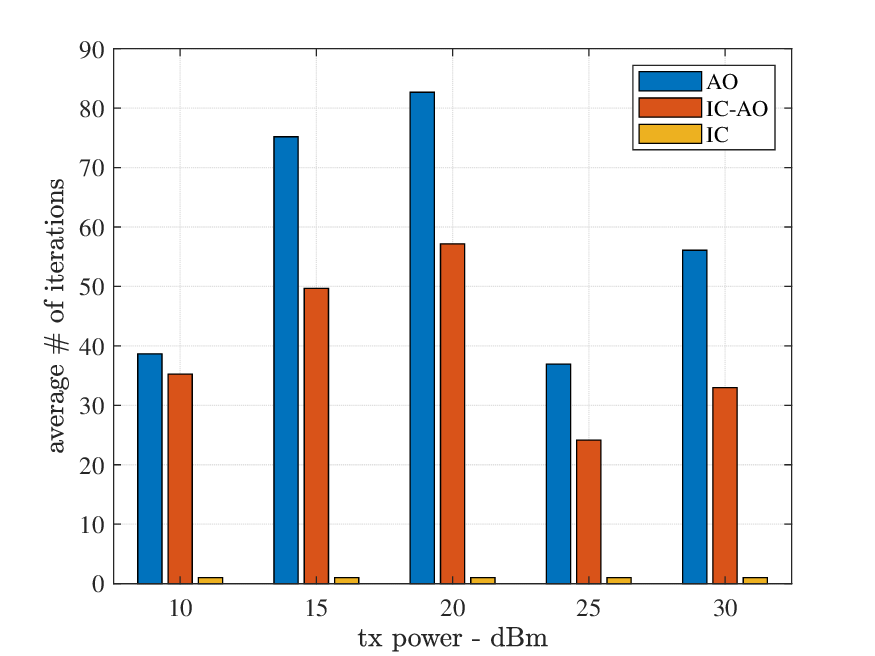}
        \caption{Computational complexity comparison}
        \label{fig_num_iter_vs_P}
    \end{subfigure}
    \caption{Performance and computational complexity comparison versus transmit power with $N = 64$ RIS elements.}
    \label{fig_performance_vs_P}
\end{figure}

\begin{figure}[t!]
    \centering
    \begin{subfigure}[t]{\linewidth}
        \centering
        \includegraphics[width=0.95\linewidth]{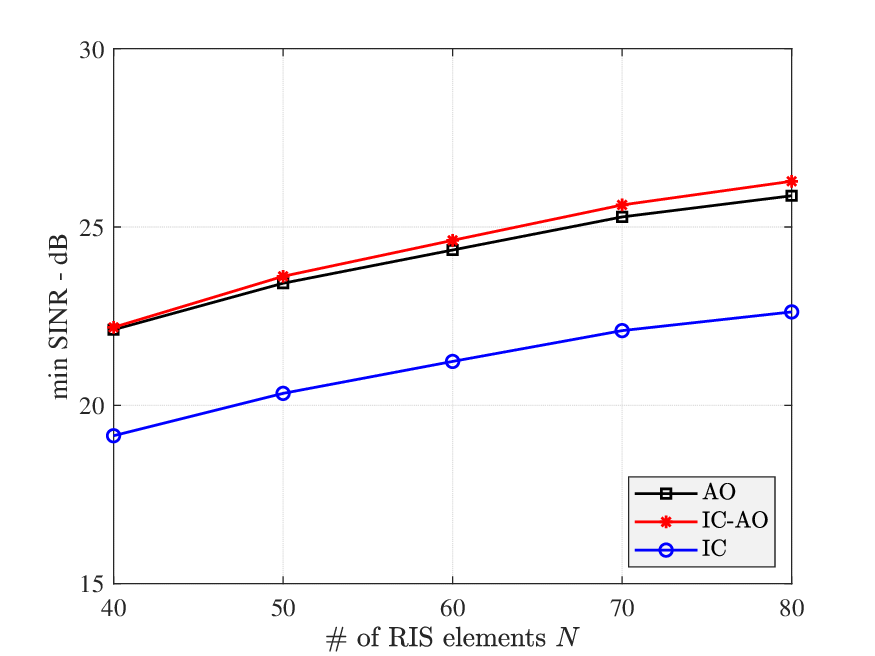}
        \caption{Minimum SINR performance}
        \label{fig_min_SINR_vs_N}
    \end{subfigure}
    
    \begin{subfigure}[t]{\linewidth}
        \centering
        \includegraphics[width=0.95\linewidth]{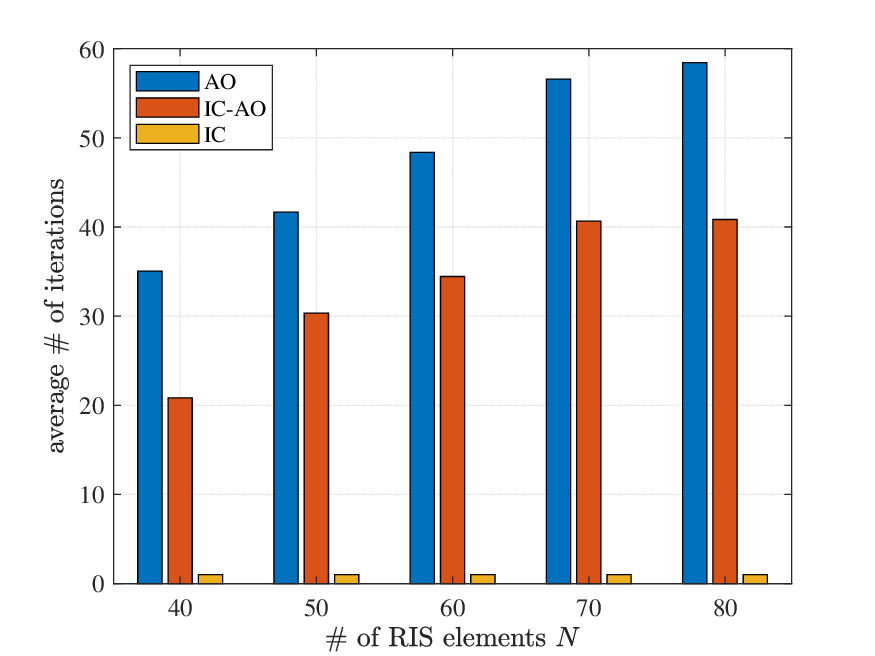}
        \caption{Computational complexity comparison}
        \label{fig_num_iter_vs_N}
    \end{subfigure}
    \caption{Performance and computational complexity comparison versus number of RIS elements $N$ at $30$-dBm transmit power.}
    \label{fig_performance_vs_N}
\end{figure}

\section{Numerical Results}
In this section, we present numerical results to show the superiority and benefits of the proposed IC and IC-AO approaches where IC-AO indicates the AO approach that uses the IC solution $\bsm{\hat{\phi}}_{\mrm{IC}}$ as an initialization. For the regular AO method, we use a random initial solution for $\bsm{\phi}$. We consider $K = 2$ users, $L = 2$ device pairs, and $M = 8$ antennas at the BS. If not specifically stated, the number of RIS elements is set to $64$. We position the BS and the RIS at the locations $(x,y)=(0,0)$ and $(x,y)=(100,30)$, respectively. The cellular users and transmit devices are randomly located within a circular area whose center is $(x,y)=(200,0)$ with a radius of $25$ m, and we also locate the receive devices randomly also within a circular area centered at $(x,y)=(50,0)$ with the same radius. The transmit power of the cellular users and the transmit devices are set to be the same. The noise power is set to $-169$ dBm/Hz and a bandwidth of $1$ MHz is assumed. The large-scale fading is modeled as $\beta = \beta_0 (d/d_0)^{-\eta}$, where $\eta$ is the path loss exponent and $d$ is the distance. We set $\beta_0 = -30$ dB as the path loss at the reference distance $d_0 = 1$m. The path loss exponent of the user-RIS, TxD-RIS, RIS-RxD, and RIS-BS channel links is set to $2.2$. However, the path loss exponent of the user-BS and TxD-BS channels is set to $4$ and the path loss exponent of the user-RxD and TxD-RxD channels are set to $5$, to model their weak propagation.

In Fig.~\ref{fig_performance_vs_P}, we show a comparison of the minimum SINR performance and computational complexity as the transmit power changes. It can be seen from Fig.~\ref{fig_min_SINR_vs_P} that the AO and IC-AO approaches give the same minimum SINRs, which are about $2.5$ to $3$ dB higher than that of the IC approach. However, the computational complexity of the IC approach is much lower than that of AO and IC-AO since the IC approach requires only one iteration while AO and IC-AO require many more iterations as seen in Fig.~\ref{fig_num_iter_vs_P}. It is also observed that although the AO and IC-AO approaches give the same SINR performance, IC-AO has a lower computational complexity since it requires a smaller number of iterations.

Next, in Fig.~\ref{fig_performance_vs_N}, we also show a comparison for the minimum SINR performance and computational complexity but with a fixed transmit power of $30$ dBm and the number of RIS elements $N$ is varied. The minimum SINRs for AO and IC-AO are still about $3$-dB higher than that of the IC approach. However, it is interesting that as the number of RIS elements $N$ increases, the IC-AO approach tends to perform better than AO even though IC-AO still requires a smaller number of iterations. This indicates that IC-AO not only converges faster than AO but also converges to better local solutions.

\begin{figure}
    \centering
    \includegraphics[width=\linewidth]{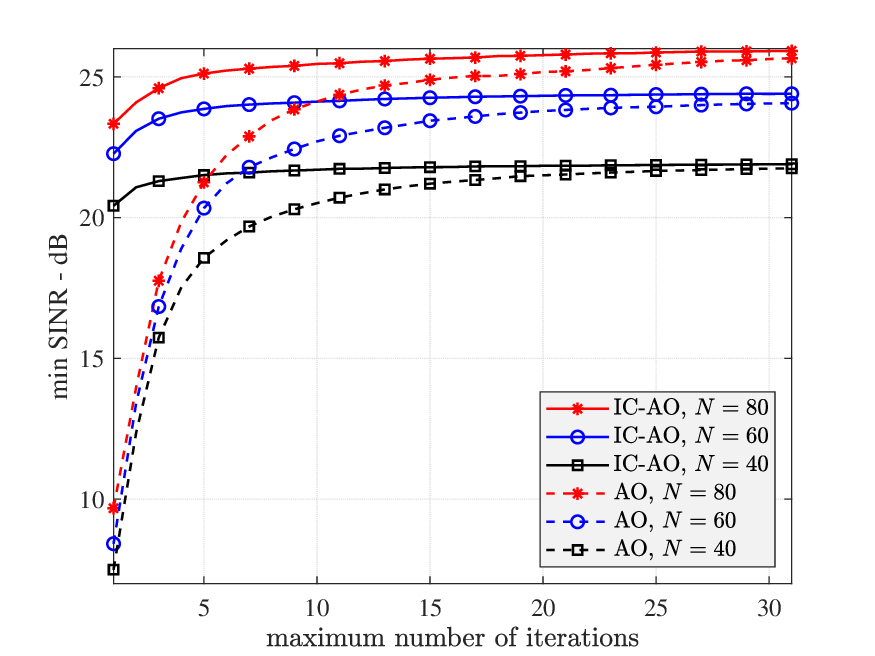}
    \caption{Minimum SINR performance versus maximum number of iterations.}
    \label{fig_performance_vs_iter}
\end{figure}

Finally, we show an SINR performance comparison between the AO and IC-AO approaches in Fig.~\ref{fig_performance_vs_iter} as the maximum number of iterations changes. It is clear that IC-AO converges much faster than AO and if we set a small maximum number of iterations, the performance of IC-AO will be significantly higher than that of AO. For example, if we set the maximum number of iterations to $5$, the IC-AO approach has already nearly converged and its minimum SINRs are about $4$-dB, $3.5$-dB, and $3$-dB higher than that of AO for $N = 80$, $60$, $40$, respectively.

\section{Conclusion}
This paper has considered a minimum SINR maximization problem in an RIS-aided joint D2D and cellular communication system where the RIS was exploited to simultaneously support the cellular users and the devices. We first presented an AO approach based on the SDR technique to solve the min-max SINR problem. We then derived a representation for the RIS phase shift vector that cancels the interference to the D2D links. Based on the IC representation, we proposed an IC algorithm whose complexity is much lower than that of the AO approach. We also showed that the IC solution can help the AO approach converge faster and achieve better performance. It was also pointed out that for the case of a single D2D pair, the proposed IC approach can be accomplished with limited feedback from the device.



\ifCLASSOPTIONcaptionsoff
  \newpage
\fi

\bibliographystyle{IEEEtran}
\bibliography{ref}

\begin{thebibliography}{10}
\providecommand{\url}[1]{#1}
\csname url@samestyle\endcsname
\providecommand{\newblock}{\relax}
\providecommand{\bibinfo}[2]{#2}
\providecommand{\BIBentrySTDinterwordspacing}{\spaceskip=0pt\relax}
\providecommand{\BIBentryALTinterwordstretchfactor}{4}
\providecommand{\BIBentryALTinterwordspacing}{\spaceskip=\fontdimen2\font plus
\BIBentryALTinterwordstretchfactor\fontdimen3\font minus
  \fontdimen4\font\relax}
\providecommand{\BIBforeignlanguage}[2]{{%
\expandafter\ifx\csname l@#1\endcsname\relax
\typeout{** WARNING: IEEEtran.bst: No hyphenation pattern has been}%
\typeout{** loaded for the language `#1'. Using the pattern for}%
\typeout{** the default language instead.}%
\else
\language=\csname l@#1\endcsname
\fi
#2}}
\providecommand{\BIBdecl}{\relax}
\BIBdecl

\bibitem{Jameel2018D2D}
F.~Jameel, Z.~Hamid, F.~Jabeen, S.~Zeadally, and M.~A. Javed, ``A survey of
  device-to-device communications: {R}esearch issues and challenges,''
  \emph{IEEE Commun. Surveys \& Tutorials}, vol.~20, no.~3, pp. 2133--2168,
  2018.

\bibitem{Gu2015Matching}
Y.~Gu, Y.~Zhang, M.~Pan, and Z.~Han, ``Matching and cheating in device to
  device communications underlying cellular networks,'' \emph{IEEE J. Select.
  Areas Commun.}, vol.~33, no.~10, pp. 2156--2166, 2015.

\bibitem{Wang2018Resource}
J.~Wang, Y.~Huang, S.~Jin, R.~Schober, X.~You, and C.~Zhao, ``Resource
  management for device-to-device communication: {A} physical layer security
  perspective,'' \emph{IEEE J. Select. Areas Commun.}, vol.~36, no.~4, pp.
  946--960, 2018.

\bibitem{Yang2016Energy}
K.~Yang, S.~Martin, C.~Xing, J.~Wu, and R.~Fan, ``Energy-efficient power
  control for device-to-device communications,'' \emph{IEEE J. Select. Areas
  Commun.}, vol.~34, no.~12, pp. 3208--3220, 2016.

\bibitem{Basar2019Wireless}
E.~Basar, M.~Di~Renzo, J.~De~Rosny, M.~Debbah, M.-S. Alouini, and R.~Zhang,
  ``Wireless communications through reconfigurable intelligent surfaces,''
  \emph{IEEE Access}, vol.~7, pp. 116\,753--116\,773, Sept. 2019.

\bibitem{Marco2020Smart}
M.~Di~Renzo, A.~Zappone, M.~Debbah, M.-S. Alouini, C.~Yuen, J.~de~Rosny, and
  S.~Tretyakov, ``Smart radio environments empowered by reconfigurable
  intelligent surfaces: {H}ow it works, state of research, and the road
  ahead,'' \emph{IEEE J. Select. Areas in Commun.}, vol.~38, no.~11, pp.
  2450--2525, Nov. 2020.

\bibitem{Tao2022Interference}
T.~Jiang and W.~Yu, ``Interference nulling using reconfigurable intelligent
  surface,'' \emph{IEEE J. Select. Areas Commun.}, vol.~40, no.~5, pp.
  1392--1406, May 2022.

\bibitem{Fangzhou2023HRIS}
F.~Wang and A.~L. Swindlehurst, ``Hybrid {RIS}-assisted interference mitigation
  for spectrum sharing,'' in \emph{Proc. IEEE Int. Conf. Acoustics, Speech and
  Signal Process.}, Rhodes Island, Greece, June 2023.

\bibitem{Fangzhou2023ARIS}
------, ``Applications of absorptive reconfigurable intelligent surfaces in
  interference mitigation and physical layer security,'' \emph{IEEE Trans.
  Wireless Commun. (Early Access)}, 2023.

\bibitem{Shah2022Statistical}
S.~W.~H. Shah, A.~N. Mian, S.~Mumtaz, A.~Al-Dulaimi, C.-L. I, and J.~Crowcroft,
  ``Statistical {Q}o{S} analysis of reconfigurable intelligent surface-assisted
  {D2D} communication,'' \emph{IEEE Trans. Veh. Technol.}, vol.~71, no.~7, pp.
  7343--7358, July 2022.

\bibitem{Yiyang2021Performance}
Y.~Ni, Y.~Liu, J.~Wang, Q.~Wang, H.~Zhao, and H.~Zhu, ``Performance analysis
  for {RIS}-assisted {D}2{D} communication under {N}akagami-$m$ fading,''
  \emph{IEEE Trans. Veh. Technol.}, vol.~70, no.~6, pp. 5865--5879, June 2021.

\bibitem{Zelin2022RIS}
Z.~Ji, Z.~Qin, and C.~G. Parini, ``Reconfigurable intelligent surface aided
  cellular networks with device-to-device users,'' \emph{IEEE Trans. Commun.},
  vol.~70, no.~3, pp. 1808--1819, Mar. 2022.

\bibitem{Fangmin2022Sumrate}
F.~Xu, Z.~Ye, H.~Cao, and Z.~Hu, ``Sum-rate optimization for {IRS}-aided {D2D}
  communication underlaying cellular networks,'' \emph{IEEE Access}, vol.~10,
  pp. 48\,499--48\,509, 2022.

\bibitem{sultana2022intelligent}
A.~Sultana and X.~Fernando, ``Intelligent reflecting surface-aided
  device-to-device communication: {A} deep reinforcement learning approach,''
  \emph{Future Internet}, vol.~14, no.~9, p. 256, 2022.

\bibitem{Yali2021RIS}
Y.~Chen, B.~Ai, H.~Zhang, Y.~Niu, L.~Song, Z.~Han, and H.~Vincent~Poor,
  ``Reconfigurable intelligent surface assisted device-to-device
  communications,'' \emph{IEEE Trans. Wireless Commun.}, vol.~20, no.~5, pp.
  2792--2804, 2021.

\bibitem{Yashuai2021Sumrate}
Y.~Cao, T.~Lv, W.~Ni, and Z.~Lin, ``Sum-rate maximization for
  multi-reconfigurable intelligent surface-assisted device-to-device
  communications,'' \emph{IEEE Trans. Commun.}, vol.~69, no.~11, pp.
  7283--7296, Nov. 2021.

\bibitem{Gang2021RIS}
G.~Yang, Y.~Liao, Y.-C. Liang, O.~Tirkkonen, G.~Wang, and X.~Zhu,
  ``Reconfigurable intelligent surface empowered device-to-device communication
  underlaying cellular networks,'' \emph{IEEE Trans. Commun.}, vol.~69, no.~11,
  pp. 7790--7805, Nov. 2021.

\bibitem{Fangshou2020Joint}
F.~Wang and H.~Li, ``Joint waveform and receiver design for co-channel hybrid
  active-passive sensing with timing uncertainty,'' \emph{IEEE Trans. Signal
  Process.}, vol.~68, pp. 466--477, 2020.

\bibitem{Luo2010SDR}
Z.~Luo, W.~Ma, A.~M. So, Y.~Ye, and S.~Zhang, ``Semidefinite relaxation of
  quadratic optimization problems,'' \emph{IEEE Signal Process. Mag.}, vol.~27,
  no.~3, pp. 20--34, May 2010.

\end{thebibliography}


%









\end{document}